# Feasibility Study of Neutron Dose for Real Time Image Guided Proton Therapy: A Monte Carlo Study

**Jin Sung Kim, Jung Suk Shin, Daehyun Kim, EunHyuk Shin, Kwangzoo Chung, Sungkoo Cho, Sung Hwan Ahn, Sanggyu Ju, Yoonsun Chung, Sang Hoon Jung, Youngyih Han**

***Department of Radiation Oncology, Samsung Medical Center, Sungkyunkwan University School of Medicine, Seoul, 135-710***

Two full rotating gantry with different nozzles (Multipurpose nozzle with MLC, Scanning Dedicated nozzle) with conventional cyclotron system is installed and under commissioning for various proton treatment options at Samsung Medical Center in Korea. The purpose of this study is to investigate neutron dose equivalent per therapeutic dose, H/D, to x-ray imaging equipment under various treatment conditions with monte carlo simulation. At first, we investigated H/D with the various modifications of the beam line devices (Scattering, Scanning, Multi-leaf collimator, Aperture, Compensator) at isocenter, 20, 40, 60 cm distance from isocenter and compared with other research groups. Next, we investigated the neutron dose at x-ray equipments used for real time imaging with various treatment conditions. Our investigation showed the 0.07 ~ 0.19 mSv/Gy at x-ray imaging equipments according to various treatment options and intestingly 50% neutron dose reduction effect of flat panel detector was observed due to multi-leaf collimator during proton scanning treatment with multipurpose nozzle. In future studies, we plan to investigate experimental measurement of neutron dose and validation of simulation data for x-ray imaging equipment with additional neutron dose reduction method.

Corresponding to Youngyih Han, Ph.D.

Department of Radiation Oncology, Samsung Medical Center

Sungkyunkwan University School of Medicine,

50 Ilwondong Gangnam-gu, Seoul, Republic of Korea, 135-710

Tel) 82-2-3410-2604

Fax) 82-2-3410-2619

E-mail: Youngyih@skku.edu

## I. INTRODUCTION

Proton radiation has totally different dosimetric characteristics from those of conventional radiation therapy [1]. X-ray radiation exhibits exponentially decaying energy deposition in tissue with increasing depth beyond a build-up region, whereas protons exhibit an increasing energy deposition with the penetration depth, with a maximum energy deposition, the Bragg peak and near the end of the proton beam range [2]. The spread-out Bragg peak, which indicates the region of maximum energy deposition, can be positioned within the target to create a conformal high-dose region. Although advanced photon radiotherapy, such as intensity-modulated radiotherapy and tomotherapy, can produce the same level of dose conformity across a tumor, proton beams have the advantage of decreasing the low-dose volume in the surrounding normal tissue [3, 4].

However, Erik reported secondary neutron dose is the critical problem with scattered proton therapy and the cancer risk is higher than IMRT treatment [5]. After this report, every proton therapy center are calculating or measuring their secondary neutron dose before their 1st treatment at different positions. [6-13]

Recently proton therapy systems are adapting advanced image guidance systems which are used and proven technology in x-ray therapy field such as CBCT and real-time tracking during treatment. Two full rotating gantry with different nozzles (Multipurpose nozzle with MLC, Scanning Dedicated nozzle) with conventional cyclotron system is installed and under commissioning for various proton treatment options at Samsung Medical Center in Korea. Since the X-ray flat panel system is not located in the proton beam nozzle, X-ray imaging during treatment to check the position of tumor or any fiducial marker is possible as shown in Fig 1. Image guided proton therapy with real-time x-ray imaging during proton treatment is the one of most advanced treatment option, but secondary neutron exposure for this imaging technique has not been studied at all.

The purpose of this study is to investigate neutron dose equivalent per therapeutic dose, H/D, to x-ray imaging equipment under various treatment conditions with monte-carlo simulation.

## II. METHOD

We have two different nozzle structures to deliver proton dose to patients: Multipurpose nozzle and PBS dedicated nozzle according to their lateral spreading method as shown in Figure 2.

### A. Multipurpose nozzle

1. Wobbling mode with multipurpose nozzle

The two x- and y- wobbling magnet generate the larger proton beam size using circular movement with fixed frequency and the enlarged proton beam pass through the scatter to make two dimensional dose distribution which has the actual field size. Proton beam with pristine bragg peak after the ridge filter can get the spread out bragg peak (SOBP) and we need several combination of ridge filter because each ridge filter has only one SOBP value. The patient aperture and multi-Leaf collimator (MLC) are a brass beam stop with a hole shaped to the outer projection of the target in the beam's eye view. The range compensator is a plastic block with material cut away in a complex shape. It is carefully aligned with the aperture and the patient's tumors and tailors the dose in depth by shifting more or less proton range depending on what part of the tumors, and upstream tissues, a particular proton ray is aimed at. So, we need these all combination of scatterer, ridge filter, MLC, compensator and aperture to deliver wobbling beam to patients.

2. Scanning mode in multipurpose nozzle

When we use scanning mode, we don't need all combinations which are used in the wobbling mode such as wobbling magnet, scatterer, ridge filter, MLC, compensator and aperture. We use only scanning magnet for proton beam delivery. However, every components, specially MLC unit are still exist in the nozzle and working as a neutron absorber if there is no collision between proton and each component.

Since recent advance scanning proton treatment using aperture to reduce the penumbra is reported, this combination for neutron dose distribution is also calculated with monte-carlo simulation

**B. PBS dedicated nozzle**

The PBS dedicated nozzle is only for scanning treatment and does not have any component to generate the neutron dose except scanning magnet and He gas duct. However we also calculated the case which use patient aperture to reduce the penumbra.

**C. MCNP simulation**

A Monte Carlo studies with the MCMPX V2.5 code were performed based on 2 different beam nozzle's geometry (MLC, compensator, aperture, ridge filter, satterer and etc) given by Sumitomo Heavy Industry and a 40 x 30 x 30 cm water phantom was simulated with 230 MeV proton beams and 10 cm SOBP using a 15 x 15 cm brass aperture. At first, we investigated H/D with the various modifications of the beam line devices (Scattering, Scanning, Multi-leaf collimator, Aperture, Compensator) at isocenter, 20, 40, 60 cm distance from isocenter and compared with other research groups. Next, we investigated the neutron dose at x-ray equipments used for real time imaging with various treatment conditions. The detail information for each treatment conditions is shown in Table 1. The schematic view of the two different proton beam delivery systems and the points whose neutron dose per absorbed dose was calculated using MCNP simulation is shown in Figure 2.

## III. RESULTS AND DISCUSSION

### A. Neutron dose distribution at isocenter axis

Table 2 lists the neutron doses received at various distance from the isocenter using six different treatment options. Our simulations presented H/D values at isocenter, 20, 40, 60 cm from isocenter for 6 different treatment options using scattering and scanning methods and showed comparable results with measured or simulated neutron dose of other proton therapy center as shown in Figure 3. We need to validate these monte carlo simulation results with actual neutron dose measurement, However we can use our monte carlo simulation for relative dose comparison to see the neutron dose effect or other systematic analysis.

Figure 4 show the neutron dose per absorbed dose at isocenter using all treatment options. Our simulation shows wobbling treatment mode (MW1, MW2) produce relatively 429% higher neutron dose than other scanning treatment mode (MS1, MS2, PS1, PS2) and there is no significant dependency on use of block or MLC for both wobbling and scanning treatment mode.

### B. Neutron dose at x-ray imaging equipment

Figure 5 show the neutron dose per proton absorbed dose at x-ray tube and flat panel detector for real-time image guided proton therapy. Our investigation showed the 0.08~0.19 mSv/Gy at real-time x-ray tube and 0.06~0.59 mSv/Gy at the flat panel detector according various treatment options and neutron dose are relatively higher with wobbling treatment mode. Although there is no significant dependence on use of the aperture, more than 50% neutron dose reduction effect of flat panel detector was interestingly observed at the flat panel detector for proton scanning treatment option with multipurpose nozzle (MS1, MS2). Since the difference between MS1, MS2 and PS1, PS2 is the presence of MLC, we think the MLC is working as a neutron absorber specially at scanning treatment mode in multipurpose nozzle

C. **Neutron dose in the nozzle component and the spectrum**

The neutron dose contributions from each component in nozzles are different. So our calculated neutron dose at several different points in different two type nozzles according the position (profile monitor, scatterer, ridge filter, dose monitor, MLC, compensator and etc.) in the nozzle using wobbling and pencil beam scanning mode is shown in Figure 6. The main neutron sources in scanning mode are only two monitors in the beam nozzle.

We used the MCNP parameterization for neutron generated angle and energy using ptrac file information to reduce Monte Carlo simulation time and its spectrum is shown in Figure 7.

## CONCLUSION

Neutron dose to the iso-center and x-ray equipment for real-time imaging under various proton treatment conditions was simulated for the first time at Samsung medical center. These are valuable reference data that can be directly compared for other imaging system for proton therapy in the literature. In future studies, we plan to investigate experimental measurement of neutron dose and validate our simulation data for x-ray imaging equipment with additional neutron dose reduction method for real time image guided proton therapy.

## ACKNOWLEDGEMENT

This research was supported by the Nuclear Safety Research Program through the Korea Radiation Safety Foundation(KORSAFe) and the Nuclear Safety and Security Commission(NSSC), Republic of Korea (Grant No.1402015), National Research Foundation of Korea (NRF) funded by The Ministry of Science, ICT & Future Planning (2012M3A9B6055201 and 2012R1A1A2042414) and also supported by Samsung Medical Center grant [GFO114008].

## REREFENCES

Table 1. Various treatment options (six options) with combination of MLC, Compensator and Block using two different nozzles in Samsung Medical Center

| Nozzle | Type | MLC | Compensator | Block | Treatment option |
|---|---|---|---|---|---|
| Multi-Purpose Nozzle | Wobbling | O | O | X | MW1 |
| | | O | O | O | MW2 |
| | Scanning | O | X | X | MS1 |
| | | O | X | O | MS2 |
| PBS Dedicated Nozzle | | X | X | X | PS1 |
| | | X | X | O | PS2 |

Table 2. The calculated neutron dose at different points according to various treatment options.

| (mSv/Gy) | MW1 | MW2 | MS1 | MS2 | PS1 | PS2 |
|---|---|---|---|---|---|---|
| X-ray Tube | 0.194 | 0.188 | 0.081 | 0.08 | 0.078 | 0.078 |
| Detector | 0.593 | 0.572 | 0.058 | 0.058 | 0.131 | 0.131 |
| Isocenter | 27.505 | 27.328 | 6.398 | 6.405 | 6.555 | 6.916 |
| 20cm | 3.463 | 3.13 | 0.392 | 0.351 | 0.492 | 0.417 |
| 40cm | 3.085 | 3.235 | 0.222 | 0.205 | 0.492 | 0.363 |
| 60cm | 1.788 | 1.764 | 0.168 | 0.173 | 0.404 | 0.363 |

Figure Captions.

Fig. 1. a) The layout of proton therapy center which has 2 conventional gantry with cyclotron in Samsung Medical Center. b) X-ray imaging system for proton therapy machine which has option for real time imaging during treatment.

Fig. 2. Schematic view of the two different proton beam delivery systems at the Samsung medical center and the points whose neutron dose per absorbed dose was calculated using MCNP simulation. A) multipurpose nozzle which has both wobbling and scanning mode and B) pencil beam dedicated nozzle

Fig. 3. A comparison of radiation dose equivalent per therapeutic dose as a function of distance from the field edge. The calculated 6 options and other measured results were shown.

Fig. 4. Simulated neutron dose equivalent per proton absorbed dose (mSv/Gy) at the isocenter for six different treatment options. The wobbling mode has relatively higher neutron dose than scanning mode.

Fig. 5. Simulated neutron dose equivalent per proton absorbed dose (mSv/Gy) at x-ray imaging equipment including x-ray tube and flat panel detector.

Fig. 6. The calculated neutron dose at several different points in different two type nozzles according the position in the nozzle using wobbling and pencil beam scanning mode.

Fig. 7. The neutron energy spectrum which is used in this study after MCNP parameterization for neutron generated angle and energy.

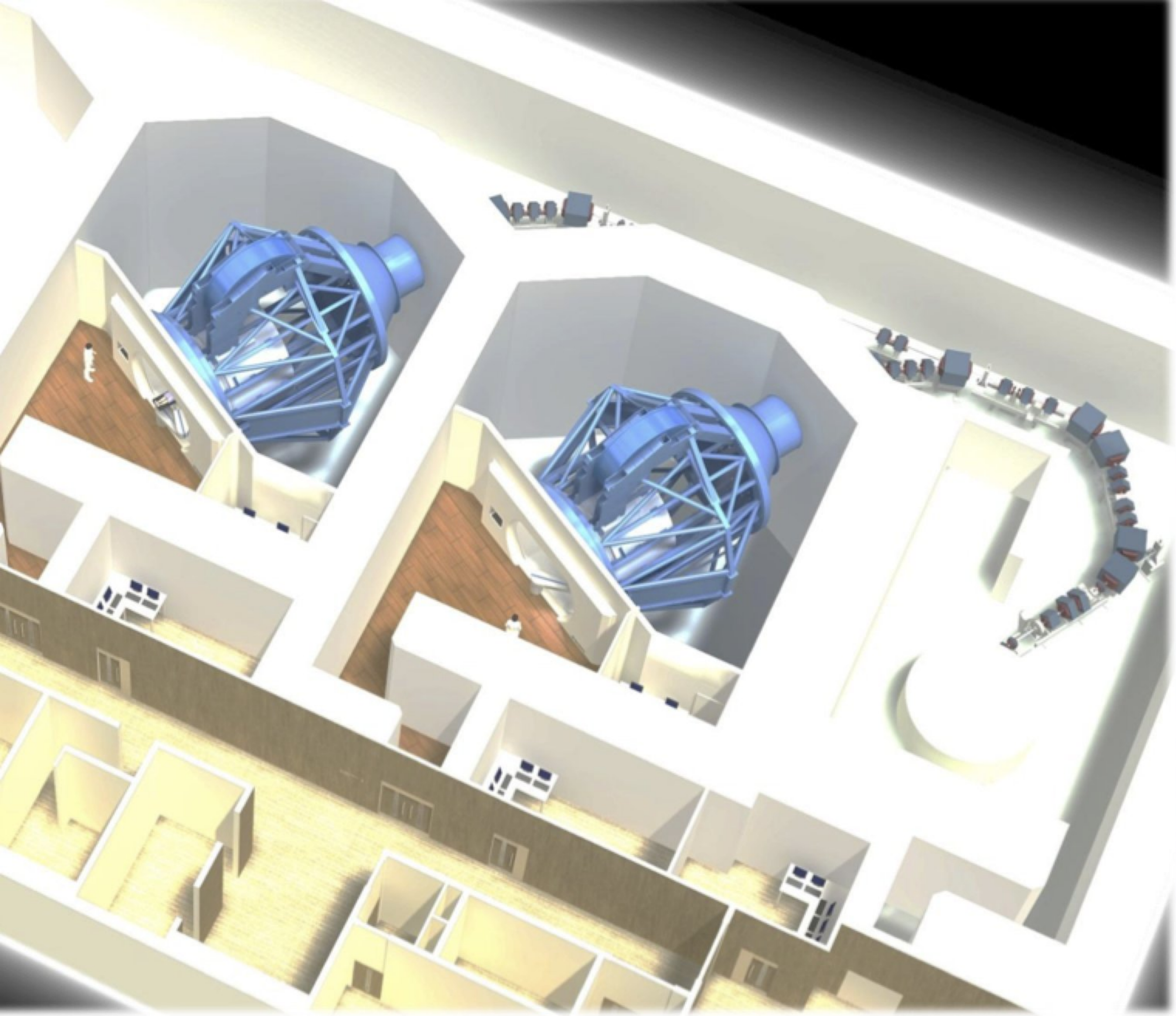

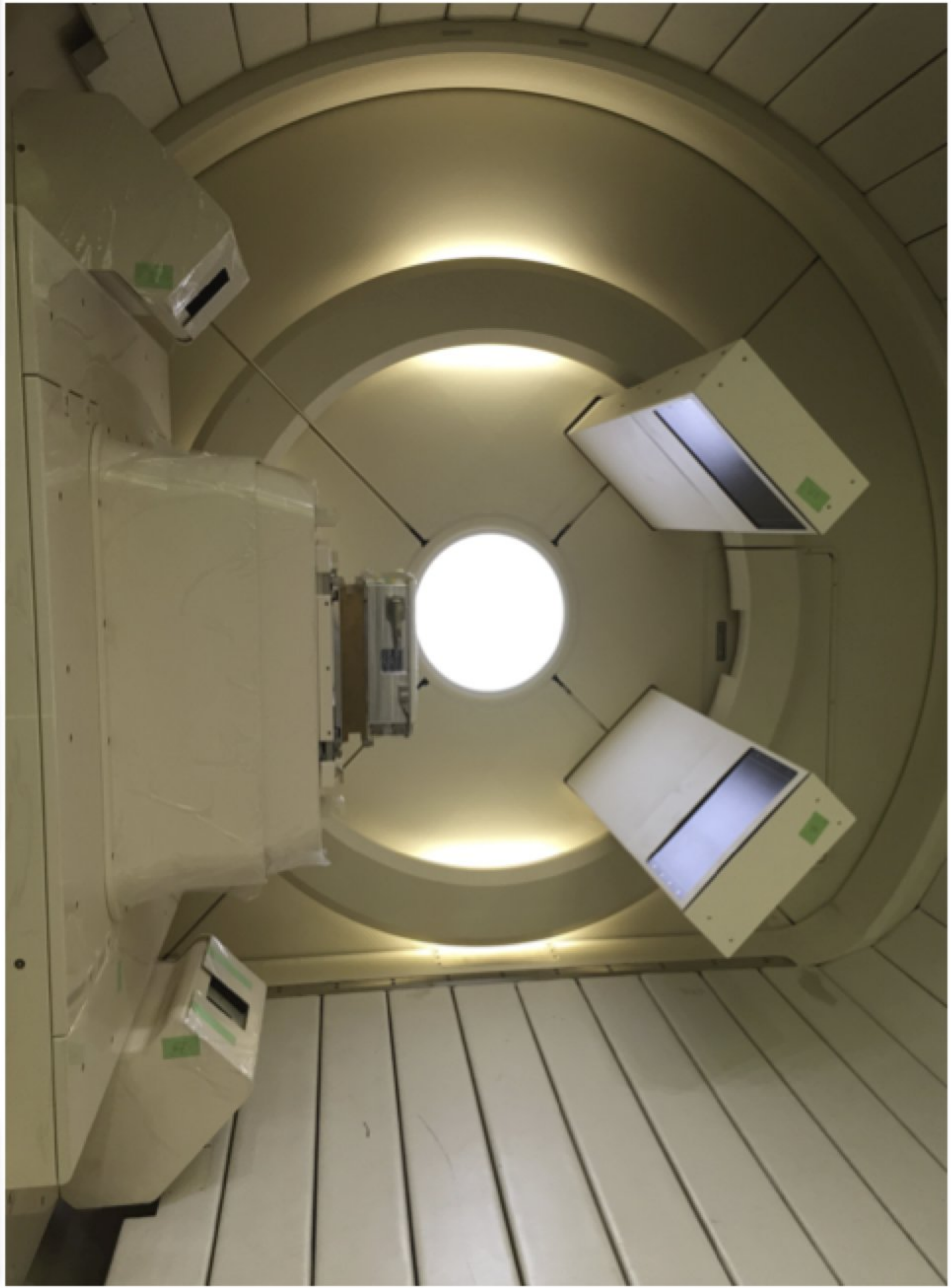

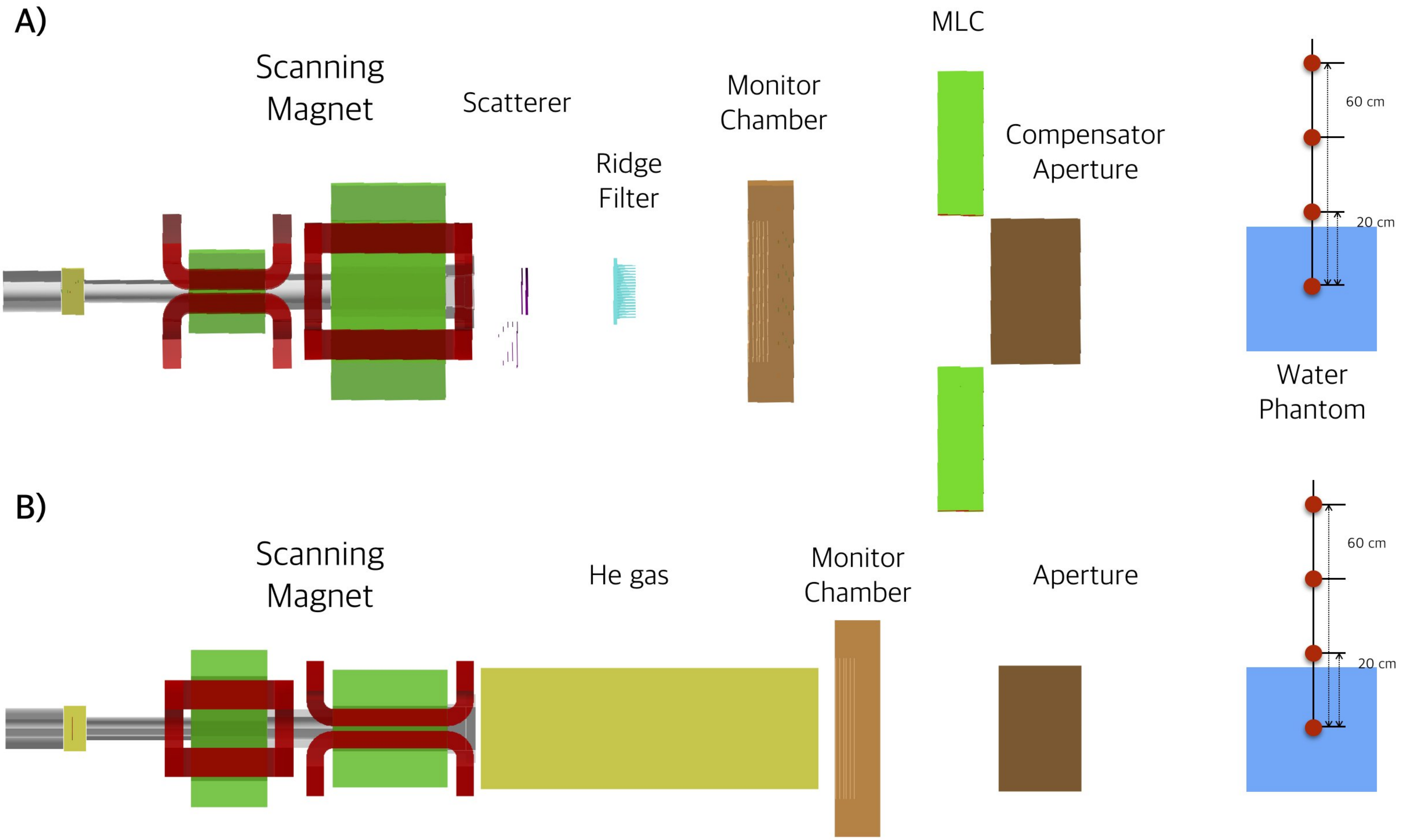

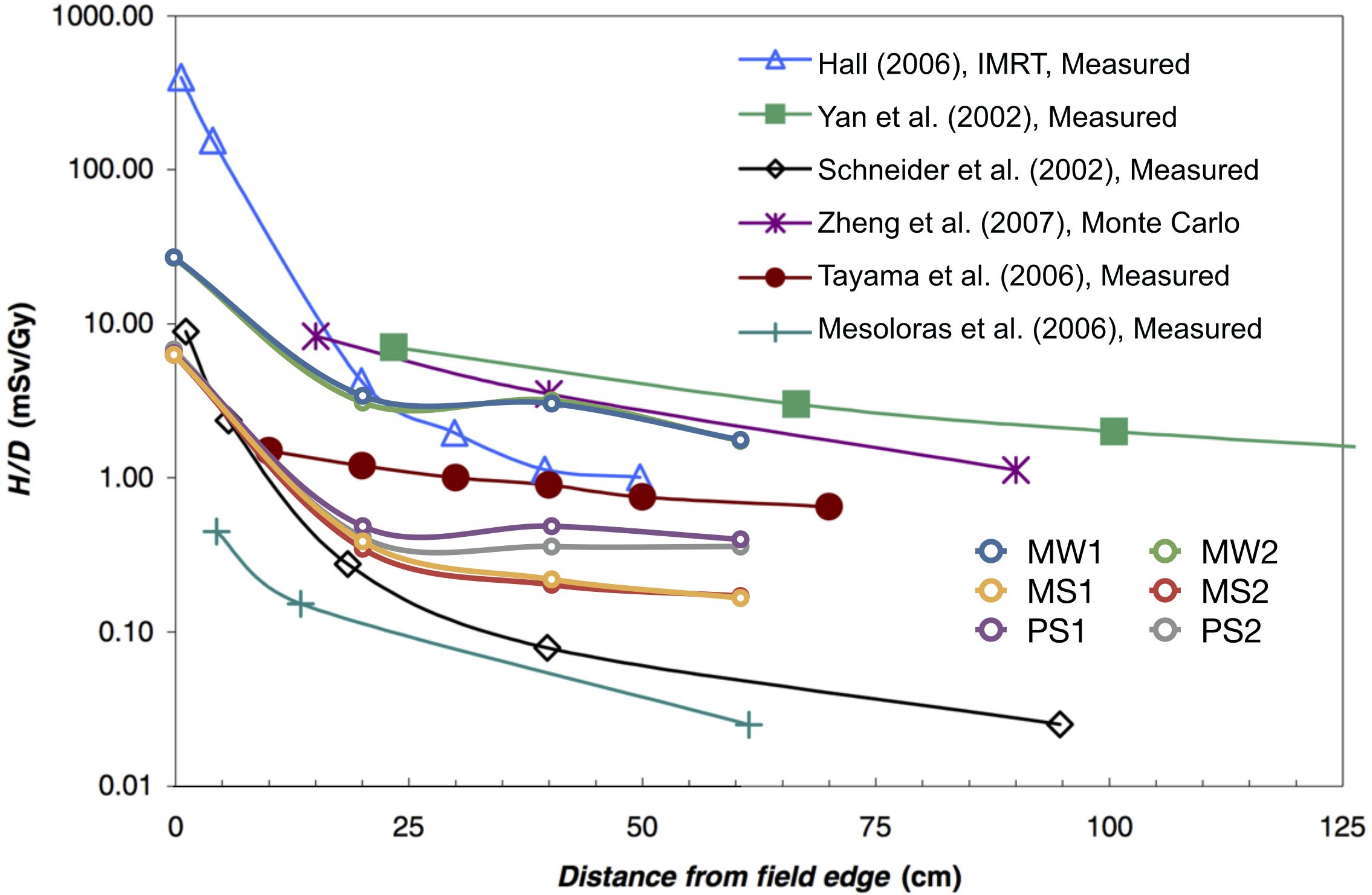

1000.00
100.00
10.00
1.00
0.10
0.01
H/D (mSv/Gy)
0
25
50
75
100
125
Distance from field edge (cm)
Hall (2006), IMRT, Measured
Yan et al. (2002), Measured
Schneider et al. (2002), Measured
Zheng et al. (2007), Monte Carlo
Tayama et al. (2006), Measured
Mesoloras et al. (2006), Measured
MW1
MW2
MS1
MS2
PS1
PS2

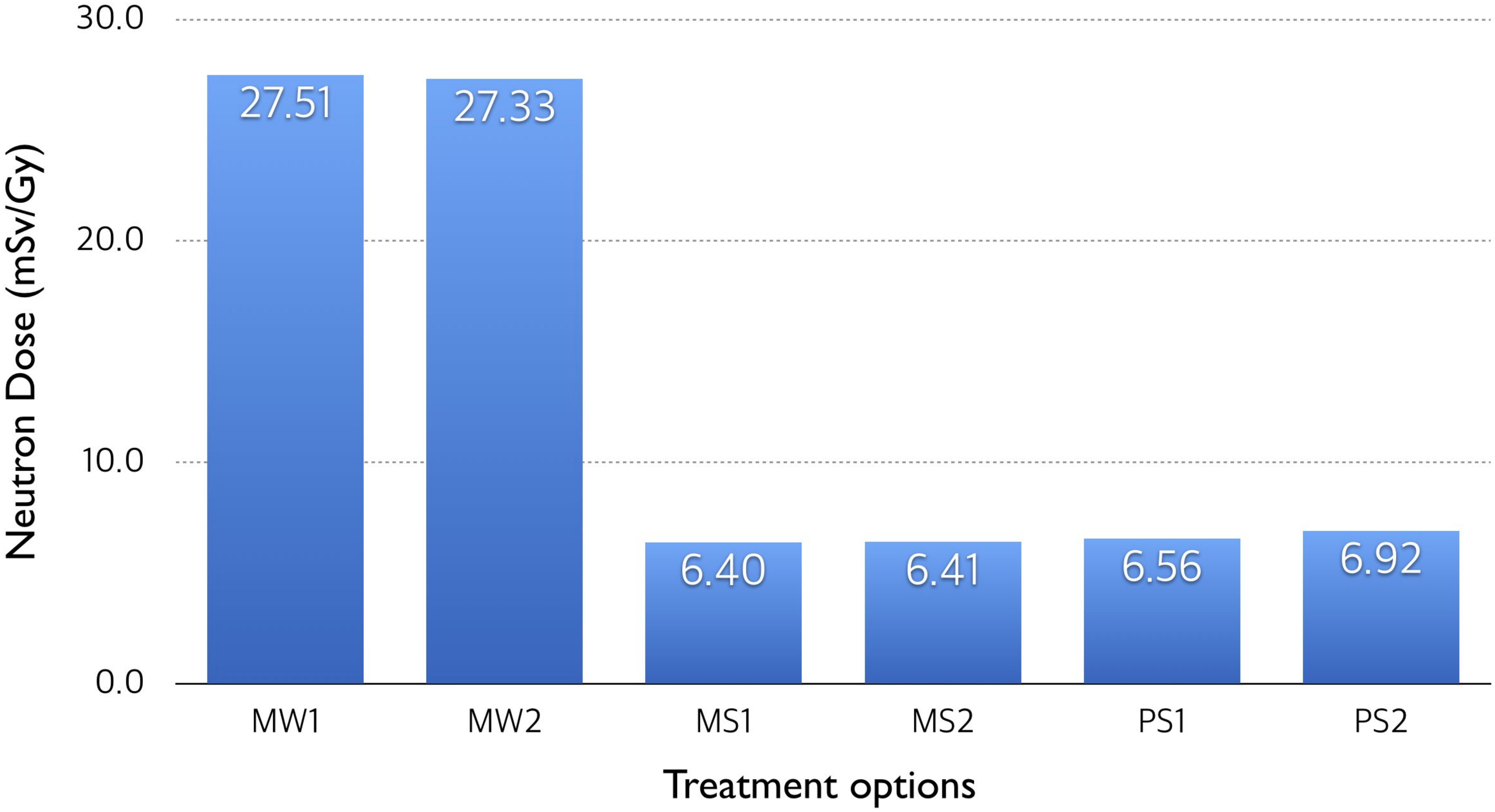

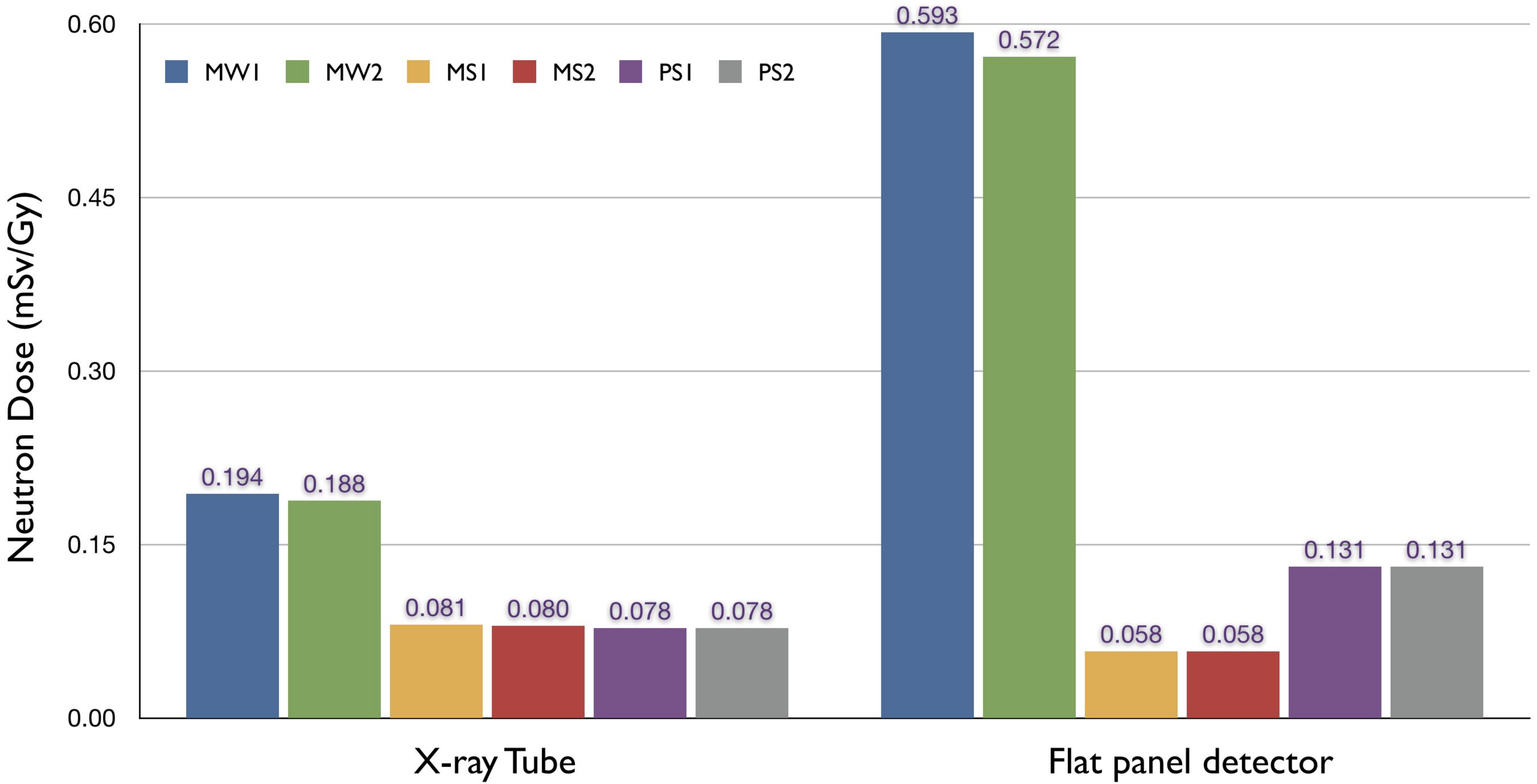

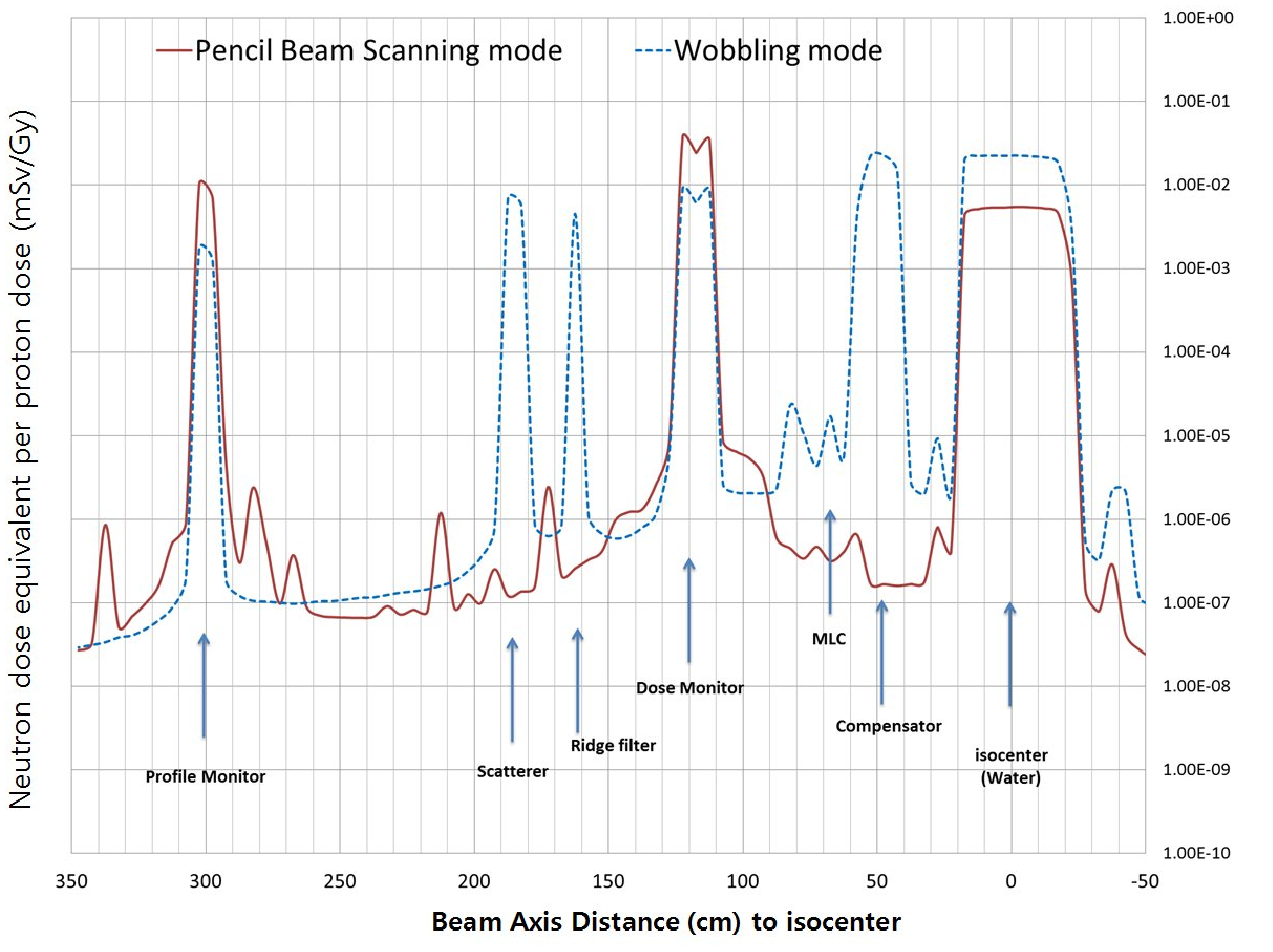

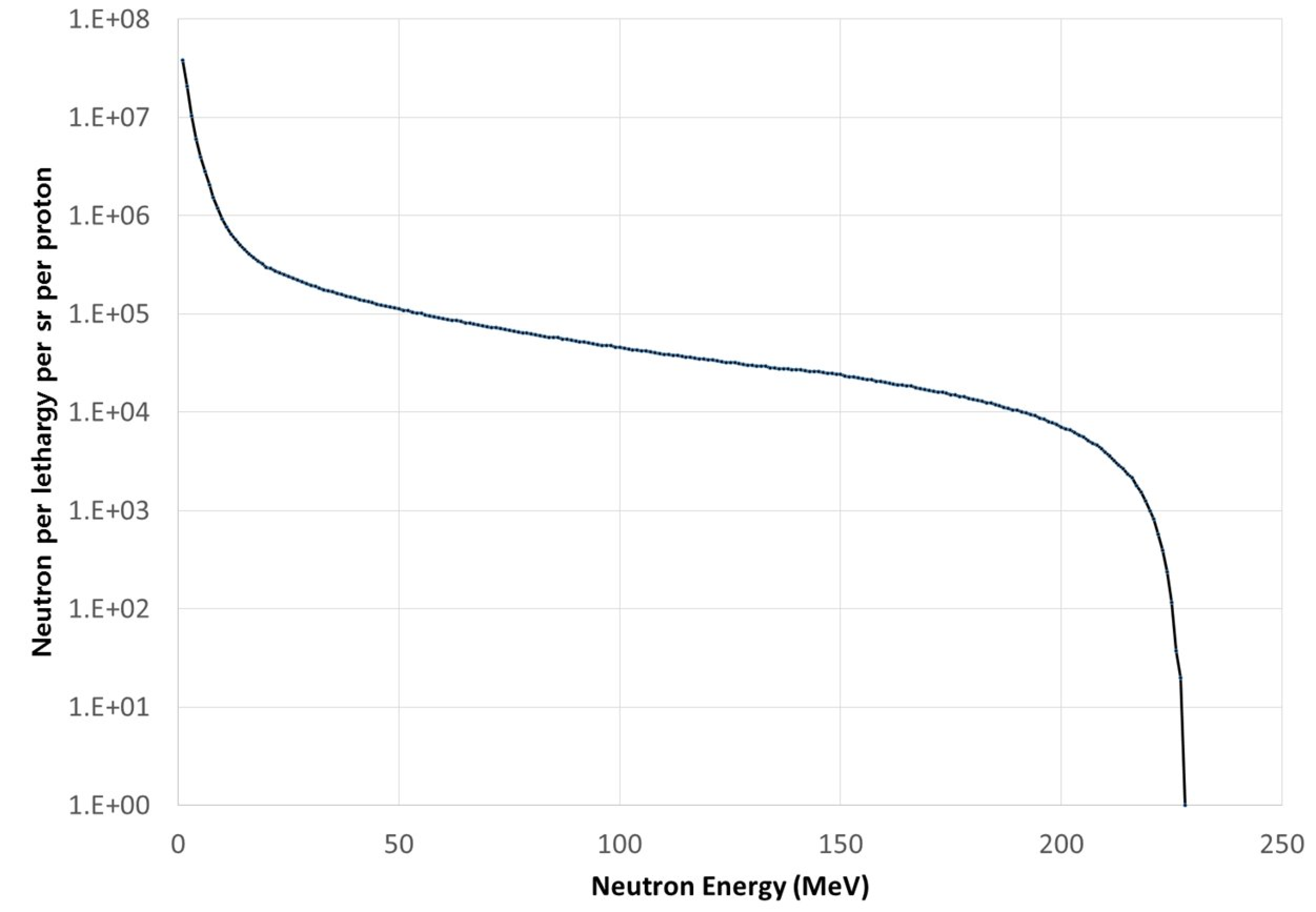